\documentclass[a4paper,12pt]{article}

\pdfoutput=1

\usepackage{inputenc}
\usepackage{authblk}
\usepackage{amsmath,amssymb}
\usepackage{graphicx}
\usepackage{epstopdf}
\usepackage{subfigure}
\usepackage{mathtools}
\usepackage{ifpdf}

\newcommand{\bes} {\begin{eqnarray*}}
\newcommand{\ees} {\end{eqnarray*}}

\newcommand{\be}{\begin{eqnarray}}
\newcommand{\ee}{\end{eqnarray}}

\newcommand{\dd} \partial

\newcommand{\ie}{{\it i.e.}\ }
\newcommand{\eg}{{\it e.g.}\ }

\title{Binder migration during drying of lithium-ion 
	battery electrodes: modelling and comparison to experiment}

\author{F. Font\thanks{Department of Physics, Universitat Polit\`{e}cnica de Catalunya, Barcelona, Spain}$\,^{\,,\dagger}$, 
B. Protas\thanks{Department of Mathematics and Statistics, McMaster University, Hamilton, Canada}, 
G. Richardson\thanks{Mathematical Sciences, University of Southampton, Hampshire, UK}, 
J. M. Foster\thanks{School of Mathematics, University of Portsmouth, Hampshire, UK}
}

\begin{document}

\maketitle

\begin{abstract}
	The drying process is a crucial step in electrode manufacture as it
	can affect the component distribution within the electrode.
	Phenomena such as binder migration can have negative effects in the
	form of poor cell performance (\eg capacity fade) or mechanical
	failure (\eg electrode delamination from the current collector). We
	present a mathematical model that tracks the evolution of the binder
	concentration in the electrode during drying. Solutions to the model
	predict that low drying rates lead to a favourable homogeneous binder
	profile across the electrode film, whereas high drying rates result
	in an unfavourable accumulation of binder near the evaporation
	surface. These results show strong qualitative agreement with
	experimental observations and provide a cogent explanation for why
	fast drying conditions result in poorly performing electrodes.
	Finally, we provide some guidelines on how the drying process could
	be optimised to offer relatively short drying times whilst
	simultaneously maintaining a roughly homogeneous binder
	distribution.
\end{abstract}


\section{Introduction} 

Lithium-ion batteries are currently used to power the vast majority of
portable electronic devices, such as cell-phones, laptops, and
tablets, and are growing in popularity for use in hybrid and electric
vehicle propulsion \cite{Lu2013}. While one of the biggest challenges
in lithium-ion battery research is to increase the energy density of
batteries, another equally important challenge is to optimize the
manufacturing process to improve long-term cycling performance and
capacity lifetime while keeping control of the manufacturing costs
\cite{Go2010,La2014,Si2015}. One particularly sensitive step in cell
production that determines the final quality of the battery pack is
the manufacturing process for the electrodes \cite{Wes15,Jai16}.

Typically, electrodes are manufactured by coating a current collector
with a slurry mixture comprised of active material (AM) particles,
conductive carbon nanoparticles, polymer binder (commonly
polyvinylidene fluoride (PVDF)) and solvent (commonly
N-Methyl-2-pyrrolidone (NMP))
\cite{Kim1999108,Lee20106049,Wes15,Jai16}. This mixture is then dried
(i.e. the solvent is evaporated) by exposure to air flow, heat and
sometimes a reduction in ambient pressure
\cite{Wes15,Jai16,Bau16,Sh2010}. The mixture preparation and coating
steps previous to drying are very important and have to be carefully
executed to ensure that electrodes are manufactured properly. For
instance, it has been shown that slurry mixtures prepared by a
multi-step process lead to a more uniform distribution of AM and
carbon particles, resulting in significantly less electrode
polarization and better cycling capability
\cite{Kim1999108,Lee20106049}. For an extensive review on mixture
preparation the reader is referred to \cite{AENM201600655}.

The most frequently used coating method in industry is slot-die
coating, in which a liquid is poured into a die that deposits the
coating liquid onto a rolling substrate belt. Coating defects such as
film instability and edge effects can occur and need to be controlled
which can, for example, be achieved by varying the coating speed and
the gap ratio \cite{Schmitt201332,Schmitt2014}. In contrast, many
research devices are manufactured by spreading the slurry on the
substrate by hand using a doctor blade. The use of NMP as a solvent is
also highly costly and replacing it with aqueous solutions would both
reduce the cost of electrode production and be more environmental
friendly \cite{WoodIII2015234}.

Drying begins once the current collector has been coated with the wet
particulate electrode mixture. The AM particles are in suspension in
the mixture whilst the binder is dissolved in the solvent. The solvent
starts evaporating from the top surface of the electrode film and the
film begins to shrink. The film once the AM particles are
in contact the film thickness stops decreasing, but evaporation continues and the pore space between
particles starts emptying. When all the solvent has been removed the
particles form a non-moving scaffold and the wet pore space has
turned into dry pore space. This being said, some recent experimental results indicate that in some circumstances pore
emptying onset even before the end of film shrinkage \cite{Jai17}.
This part of the process has been subject of intense experimental
research in recent years
\cite{Wes15,Jai16,Bau16,Jai17,Hagiwara2014,Jai17b,Muller20171}. There
is now a consensus that changes to the drying process parameters
(temperature, air-flow and pressure) significantly affect the final
electrode microstructure and, therefore, the electrochemical and
mechanical properties of the resulting battery electrode. It has been
observed by several experimental groups that high temperatures and
drying rates lead to an accumulation of binder at the film evaporation
surface and a corresponding depletion at the film-substrate interface
\cite{Wes15,Bau16,Hagiwara2014,Muller20171}. The consequence of binder
inhomogeneities include lower adhesion of the electrode to the current
collector \cite{Wes15,Jai16,Bau16}, increased electrical resistivity
\cite{Wes15} and decreased cell capacity \cite{Jai16}. Chou {\it et
	al} \cite{Chou14} conclude that even though binder makes up only a
small fraction of the electrode composition, it plays a very important
role in the cycling stability and rate capability of the electrode.

Investigation depletion and/or accumulation of binder in different regions of
electrodes has been a matter of experimental investigation
\cite{Jai16,Bau16,Jai17b,Muller20171}. In Jaiser {\it et al}
\cite{Jai16} a ``top-down'' film consolidation process is suggested, in which a dense layer, or `crust', appears on
the drying surface and grows down until it reaches the
substrate interface. However, a follow-up study seems to indicate that
film shrinkage may occur in a more homogeneous fashion \cite{Jai17b}.
In both cases, it was found that removal of the solvent from the film
surface causes enrichment of binder in the upper regions and that this
upward transport cannot be compensated by diffusion if the drying
rates are high. The effect of the drying temperature on the drying
process is discussed in \cite{Bau16}. These results show that higher
temperatures negatively influence electrode adhesion to the current
collector due to binder depletion at the electrode-collector
interface. The detrimental effect of high drying temperatures has been
recently confirmed via energy dispersive x-ray spectroscopy in
\cite{Muller20171}.

Theoretical models detailing the physical mechanisms governing the
drying of a single-component colloidal suspension were studied in
\cite{Sty11} while drying of polymer solutions was investigated in
\cite{Hen14,Hen16}. However, to the best of our knowledge, the process
of drying suspensions composed of colloidal particles and dissolved
binders has not been tackled before. The aims of this work are to: (i)
provide a mathematical model for such a situation, (ii) to compare the predictions of this model with experimental results, and (iii) use the model to suggest strategies to optimise the drying process. The rest of the
paper is organized as follows. In the next section we formulate and
solve a simple model for mass transport within an electrode when the
colloidal suspension of the AM particles is stable and the particles
remain separated and distributed homogeneously until full
consolidation has occurred. We then formulate the model for the
transport of binder through the drying film and present estimates for
the parameters in the model. In the next section, \S\ref{results}, we
present both numerical and approximate (asymptotic) solutions of the
model for different drying rates and protocols. Finally, in
\S\ref{conc} we draw our conclusions.

\begin{figure}
	\centering 
	\includegraphics[width=0.65\textwidth]{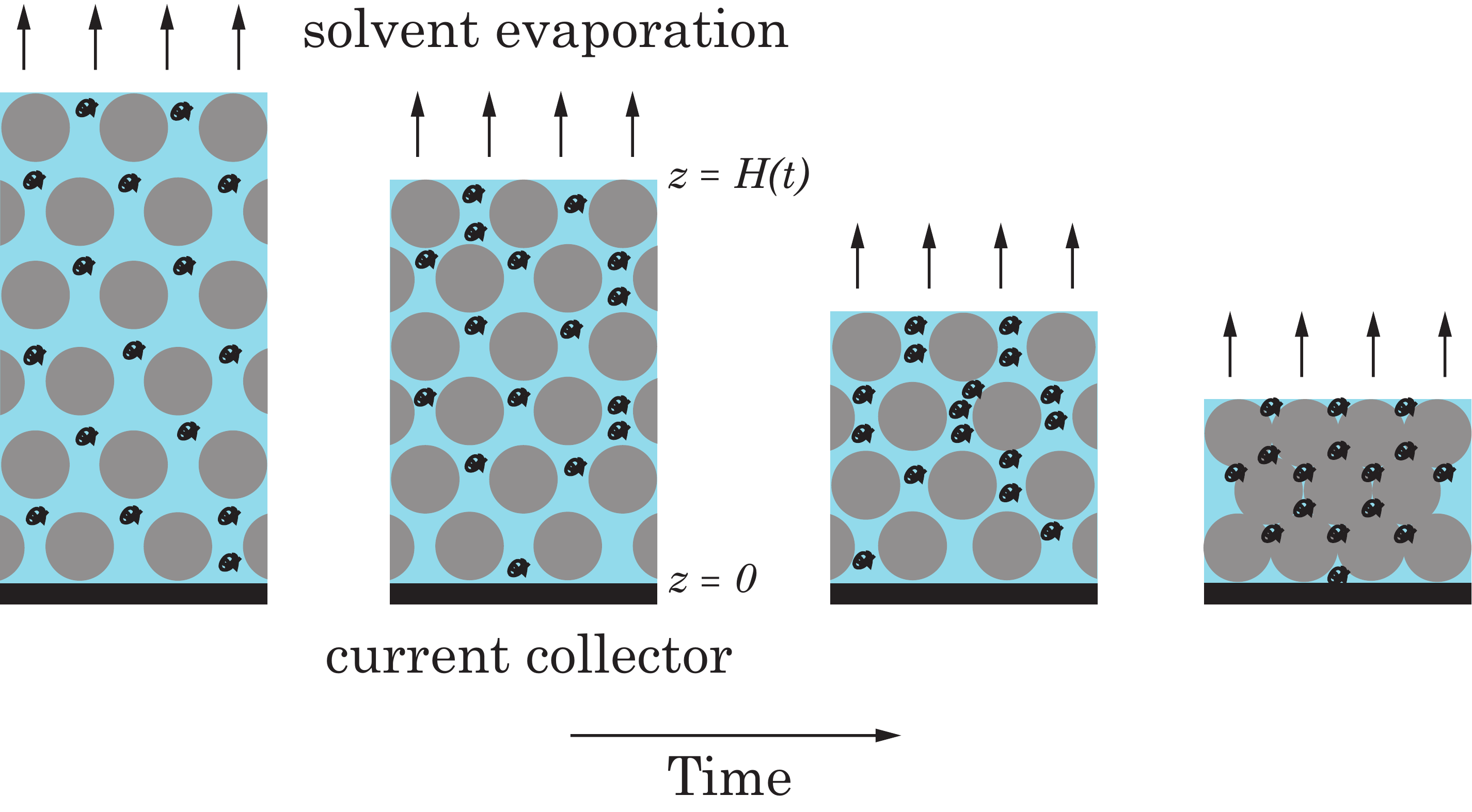}
	\caption{Illustration of the drying process of an electrode film
		sitting on top of a current collector. The blue background
		represents the solvent (which evaporates from the top surface of the
		film), and the black and grey particles represent the polymer binder and AM particles respectively.}
	\label{pic1}
\end{figure}

\section{Problem formulation}

We formulate a one-dimensional model in which mass transfer occurs
only in the $z$-direction (perpendicular to the substrate).  All model equations are defined for $z \in [0,H(t)]$,
where $H(t)$ is the time-dependent position of the top of the
electrode film, see Figure \ref{pic1}.  The assumption that the model
is one-dimensional is justified by the fact that the electrode film is
slender, \ie its lateral extent is much larger than its thickness
(height).  We will track two material phases: a liquid phase
(comprised of both the solvent and dissolved binder), with volume
fraction $\phi_{\text{l}}$, and a solid phase (AM particles), with
volume fraction $\phi_{\text{s}}$. We begin by presenting some scaling arguments that aim to identify the important effects at play during drying.

First, we characterise the typical time required for an AM particle to
settle through the NMP film. Using the material properties of the NMP
solvent and the graphite active material collected in Table
\ref{table1} together with an estimate of the typical viscosity of NMP
at 348K of $\eta=1.65$mPa\,s \cite{Hen04}, we can estimate a time
scale for sedimentation of a small 5$\mu$m graphite AM particle
through a 120$\mu$m NMP film of around 10s, which is very much faster
than the typical drying time of around 1 minute (note that larger
particles sediment even more quickly). It is therefore apparent that
the flows resulting from the drying process cannot lead to significant
AM particle redistribution within the film (the gravitational buoyancy
forces always dominate the drag forces from the drying flows).
Furthermore, experimental evidence presented in \cite{Jai17b}
indicates that AM particle distribution within the film is uniform
throughout the drying process which, in turn, suggests that there is
another physical force dominant over both buoyancy and drag forces.
This we believe to be a repulsive colloidal force resulting from
charge on the AM particles. Henceforth, in line with data presented in
\cite{Jai17b}, we assume that AM particles are always uniformly
distributed within the film.

It remains to specify a model for the transport of the PVDF binder
particles. These particles are typically very small, with a
hydrodynamic radius of around 15nm \cite{Park07}, and therefore are
largely unaffected by gravitational buoyancy effects over the drying time of
the film.  Nevertheless, it seems conceivable that they may be
affected by colloidal forces. In order to counter this hypothesis, we
note firstly that PVDF is an inert polymer, and so unlikely to be
charged, and secondly that experiments conducted in \cite{Jai16} show
that distribution of binder is strongly dependent on drying rate,
refuting the hypothesis that PVDF forms a stabilised colloidal
suspension in NMP. We therefore model transport of PVDF within the NMP
solvent by an advection diffusion equation in order to include both
the effect of the drying flows and thermal diffusion on the motion of
PVDF particles. Changes in the macroscopic (effective) diffusivity of
the binder (caused by changes in the volume fractions of the different
phases through the film) will be accounted for via use of the
Bruggemann approximation \cite{Brug}. Finally, we note that crystallisation only begins to occur when the mass fraction of the binder reaches 77 wt\% at 60$^o$C (or larger for higher temperatures)
\cite{Ma2011}. The parameters in Table \ref{table1} indicate that these concentrations will likely not occur during the stages of the drying process under consideration here, see the discussion at the end of \S\ref{sec:bintrans}. Crystallisation effects are therefore neglected.

The model will be formulated in two parts. First, we will construct
the model for the mass transport of the liquid phase (dissolved binder
and solvent) and solid phase (AM particles). Then, we will obtain the
equations describing the advection and diffusion of binder through the
moving solvent. We can adopt this approach because the volume fraction
of the binder is so small (typically only 0.01--0.05
\cite{Lee20106049,Wes15,Jai16,Jai17b}) that, to a good approximation,
the binder concentration does not influence the mass transport of the
liquid and solid phases. However, the advection-diffusion model for
the binder concentration can only be solved with knowledge of the
solvent flow. This suggests a two-step solution process in which we
first solve for particle and solvent flow and then use the results
obtained as input to the binder transport model. Finally, we solve
this model to obtain the evolution of the binder concentration.

\subsection{\label{masstrans}Mass transport}

Conservation equations for the mass of the solid (AM particles) and
liquid (binder and solvent) phases expressed in terms of their
respective volume fractions have the form
\begin{subequations} 
	\label{one}
	\begin{align}
	\frac{\partial \phi_{\text{s}}}{\partial t} + \frac{\partial F_{\text{s}}}{\partial z} & = 0, 
	\label{onea} \\ 
	\frac{\partial \phi_{\text{l}}}{\partial t} + \frac{\partial F_{\text{l}}}{\partial z} & = 0,
	\label{oneb}
	\end{align}
\end{subequations}
where $F_{\text{l}}$ and $F_{\text{s}}$ are the volume fraction fluxes
(with dimensions of m/s) of the solid and liquid phases, respectively.
In turn, these are related to the volume averaged velocities in each
phase via
\begin{subequations} 
	\label{two}
	\begin{align}
	F_{\text{s}} & = v_{\text{s}} \phi_{\text{s}},  \label{twoa} \\
	F_{\text{l}} & = v_{\text{l}} \phi_{\text{l}}.  \label{twob}
	\end{align}
\end{subequations}
Since there are only two phases (liquid and solid) their volume
fractions sum to one, \ie
\begin{equation} \label{six}
\phi_{\text{s}} + \phi_{\text{l}} = 1.
\end{equation}
As drying proceeds and liquid is removed from the slurry, the
resulting dynamics will depend on the stability of the homogeneous
state of the suspension. This is described by the widely accepted
Derjaguin-Landau-Verwey-Overbeek (DLVO) theory \cite{Horinek2014} in
which the force between two spherical particles is comprised of two
contributing parts; relatively long-range Coulombic repulsion, that is
screened by a counterion cloud, and short-range van der Waals
attraction. If the repulsive electrostatic barrier is weak, then as
solvent is removed and particles are forced into closer vicinity, they
will quickly begin to aggregate forming a crust \cite{Goe}. In
contrast, if the repulsive barrier is strong (as it is here), then particles will
remain well-separated until the volume of the film has been reduced so
much that they are forced into contact. In the latter case, the
suspension is stable and the solid phase is forced to be homogeneously
distributed throughout the electrode film (as seen in \cite{Jai17b}),
\ie we have \be \phi_{\text{s}}=\phi_{\text{s}}(t)\,, \ee.

The equations above are solved subject to no-flux boundary conditions
on the current collector, namely,
\begin{subequations} 
	\label{three}
	\begin{align}
	F_{\text{s}}|_{z=0} & = 0, \label{threea} \\
	F_{\text{l}}|_{z=0} & = 0, \label{threeb}
	\end{align}
\end{subequations}
and the following flux conditions on the evaporation surface $z=H(t)$:
\begin{align} 
\label{four1} \frac{D_{\text{s}}}{Dt} (z-H)\Big{|}_{z=H(t)}=0 \qquad \Longrightarrow \qquad  -\dot{H} + & v_{\text{s}}\Big{|}_{z=H(t)} = 0, \\
\label{five1} \frac{D_{\text{l}}}{Dt} (z-H)\Big{|}_{z=H(t)} = \frac{\gamma}{\phi_{\text{l}}}\Big{|}_{z=H(t)} \qquad\Longrightarrow \qquad -\dot{H} + & v_{\text{l}}\Big{|}_{z=H(t)} = \frac{\gamma}{\phi_{\text{l}}} \Big{|}_{z=H(t)},
\end{align}
which represent zero-flux of the solid phase and an evaporation flux
$\gamma$ of the liquid phase, respectively, through the surface
$z=H(t)$. In (\ref{four1}) and (\ref{five1}) the operators
$D_{\text{s}}/Dt$ and $D_{\text{l}}/Dt$ are material derivatives taken
with respect to the solid and liquid velocities, respectively, and a dot indicates a derivative with respect to time. At the
beginning of the drying process we assume that the two phases are well
mixed and are present in the following proportions
\begin{subequations} 
	\label{seven}
	\begin{align} 
	\phi_{\text{s}} |_{t=0} & = \phi_{\text{s}}^0 ,  \label{sevena} \\
	\phi_{\text{l}} |_{t=0} & = 1-\phi_{\text{s}}^0, \label{sevenb}
	\end{align}
\end{subequations}
whilst the film is taken to have initial thickness
\begin{equation} \label{farts}
H|_{t=0} = H_0.
\end{equation}
Validity of the model terminates at the time $t_{\text{end}}$ when the
AM particles are consolidated (\ie they make direct contact with each
other) and the liquid surface begins to intrude into the scaffold
formed by the electrode particles. We define the solid volume fraction
at this fully consolidated stage to be $\phi_{\text{s}}^{\text{max}} =
\phi_{\text{s}}(t_{\text{end}})$ and model solutions will be
terminated when this state is reached.

Summing equations \eqref{onea}--\eqref{oneb}, using \eqref{six},
integrating with respect to $z$ and imposing the boundary conditions
\eqref{three} reveals that \be \label{nonetflux}
F_{\text{l}}+F_{\text{s}} = 0.  \ee Substituting the above into the
sum of the boundary conditions \eqref{four1} and \eqref{five1}, and
using \eqref{two} and \eqref{six}, gives the following evolution
equation for the position of the top surface of the film \be
\label{shsh} \dot{H} = -\gamma.  \ee This result can readily be
interpreted as global mass conservation throughout the film.

As illustrated in Figure \ref{pic1}, the solid volume fraction is
space-independent because of strong repulsion between AM particles.
Thus, equation \eqref{onea} can be integrated with respect to $z$ and
\eqref{threea} imposed to give \be \label{rufflesniff} F_{\text{s}} =
-\dot{\phi}_{\text{s}} z.  \ee Eliminating $F_{\text{s}}$ from the
above in favour of $v_{\text{s}}$ using \eqref{twoa} and using the
boundary condition (\ref{four1}) gives $d/dt(H\phi_{\text{s}})=0$.
This can be integrated and the initial conditions \eqref{sevena} and
(\ref{farts}) imposed to give $\phi_{\text{s}} =
H_0\phi^0_{\text{s}}/H$. Back substitution of this result into
(\ref{rufflesniff}), then using (\ref{nonetflux}) and (\ref{shsh})
gives the following expressions for the volume fractions and
volume-averaged fluxes \be \label{mmaatt}
\begin{split} 
	\phi_{\text{s}} = \frac{\phi_{\text{s}}^0 H_0}{H}, \quad \phi_{\text{l}} = 1-\frac{\phi_{\text{s}}^0 H_0}{H}, \quad F_{\text{s}} = -\frac{\gamma \phi_{\text{s}}^0 H_0}{H^2} z, \quad F_{\text{l}} = \frac{\gamma \phi_{\text{s}}^0 H_0}{H^2} z\,. 
\end{split}
\ee

\subsection{\label{sec:bintrans}Binder transport}

The polymer binder is distributed within the liquid phase only, thus,
a volume-averaged continuity equation describing the concentration $c$ of
dissolved binder in the solvent is 
\be \label{eq_bin}
\frac{\dd}{\dd t} \left( \phi_{\text{l}} c \right) + \frac{\dd J}{\dd
	z} =0, \qquad J = F_{\text{l}} c - D_{\text{eff}} \frac{\dd c}{\dd
	z}, 
\ee 
where $J$ is the volume-averaged mass flux of dissolved
binder. This flux is composed of two parts: an advective part with the
volume-averaged velocity of the solvent and a diffusive part with an
``effective'' diffusion coefficient $D_{\text{eff}}$. We estimate this
effective diffusivity using the Bruggemann approximation which assumes
that $D_{\text{eff}} = D \phi_{\text{l}}^{3/2}$, where $D$ is the
diffusivity of the binder in the solvent \cite{Brug}. So, the value of
$D_{\text{eff}}$ changes during the drying process as 
$\phi_{\text{l}}$ varies according to \eqref{mmaatt}.

Suitable boundary conditions on (\ref{eq_bin}) require that there is
zero flux of binder through both the current collector and the free
surface $z=H(t)$. We therefore have 
\be \label{bc_bin} J|_{z=0} = 0,
\qquad J|_{z=H(t)} = \phi_{\text{l}} c \frac{dH}{dt}.  
\ee 
One can verify using the boundary conditions (\ref{bc_bin}) and
Leibniz integral rule on (\ref{eq_bin}) that the total amount of
binder in the film $\int_0^{H(t)} c\phi_{\text{l}}\, dz$ is
conserved throughout the drying process. We assume that initially the binder
is homogeneously distributed throughout the solvent. Thus, a suitable
initial condition to close (\ref{eq_bin}) is \be \label{init_bin}
c|_{t=0}=c_{0}.  \ee It should be noted that the model presented here
is only valid until binder concentrations get large enough that the
PVDF begins to crystallize out of solution. At 60\,$^o$C the mass
fraction for crystallization of PVDF from NMP is 77 wt\% and this
value increases with temperature \cite{Ma2011}. As a reference for the
parameter estimation we will take values from \cite{Jai16,Jai17b}
where drying was performed at 76.5\,$^o$C, so we expect mass fraction
for crystallization to be slightly above 77 wt\%. As we will show
later, in \S\ref{viable}, these concentrations are only achieved under
extremely aggressive drying rates.

\subsection{\label{sec_param}Parameter estimates}

We calibrate simulations using the data provided in
\cite{Jai16,Jai17b} and summarised here in Table \ref{table1}. We find
that the initial volume fraction of the solid electrode particles is
\begin{align}
\phi_{\text{s}}^{0} = \frac{\omega_{\text{s}}^{0}}{\rho_{\text{s}}}\left(\frac{\omega_{\text{s}}^{0}}{\rho_{\text{s}}}
+\frac{\omega_{\text{b}}^{0}}{\rho_{\text{b}}}+\frac{\omega_{\text{NMP}}^{0}}{\rho_{\text{NMP}}}\right)^{-1} \approx 0.2792\,.
\end{align} 
where the subscripts ``s'', ``b'' and ``NMP'' indicate solid electrode
particles, binder and solvent, respectively. To compute the final (and
maximal) value of the solid volume fraction we also make use of the
measured porosity of the dried electrode film $p = 0.46$ \cite{Jai17b}
and find that
\begin{align}
\phi_{\text{s}}^{\text{max}} = (1-p)\frac{\omega_{\text{s}}^{\text{max}}}{\rho_{\text{s}}}\left(\frac{\omega_{\text{s}}^{\text{max}}}{\rho_{\text{s}}}
+\frac{\omega_{\text{b}}^{\text{max}}}{\rho_{\text{b}}}+\frac{\omega_{\text{NMP}}^{0}}{\rho_{\text{NMP}}}\right)^{-1} \approx 0.5032\,. 
\end{align} 
The initial concentration of binder in the solvent and film thickness
are
\begin{equation}
c_{0}=\frac{\tilde{c}_{0}}{(1-\phi_{\text{s}}^{0})} \approx 56.95\,\text{kg/}\text{m}^3, \quad H_0 \approx 114\,\mu\text{m},
\end{equation}
where $\tilde{c}_{0} = \rho w_{\text{b}}^{0} = 41.05$\,kg/m$^{3}$ with
$\rho = \rho_{\text{s}} w_{\text{s}}^{0} + \rho_{\text{b}}
w_{\text{b}}^{0} + \rho_{\text{NMP}} w_{\text{NMP}}^{0}$ is the
initial concentration of binder in the film \cite{Jai16,Jai17,Jai17b}.
In the subsequent section we will consider how the dynamics change
with varying drying rate. Nonetheless, we note that a typical value of
the mass flux across the evaporation surface is $q_{\text{s}} =
1.19$\,g\,m$^{-2}$\,s$^{-1}$ \cite{Jai16,Jai17,Jai17b}, so that a
typical value for $\gamma$ is given by
\begin{equation}
\gamma = \frac{q_{\text{s}}}{\rho_{\text{NMP}}} \approx 1.16\,\mu \mbox{m s}^{-1}.
\end{equation}  

As a reference for the diffusion coefficient we will use $D
=1.14\cdot10^{-10}$\,m$^2$/s which is obtained, as described in the
supplementary information, by qualitatively matching the
solution of our model to the experimental results presented in
\cite{Jai16}. If we consider the viscosity of NMP $\eta =
1.65$\,mPa\,s and a temperature $T=348$\,K, the estimate for the
hydrodynamic radius of a PVDF particle using the Stokes-Einstein
relation is $R_H = k_B T/6\pi\eta D = 2.48$\,nm, which is close to the
hydrodynamic radius found experimentally for PVDF chains in
PVDF/Propylene-carbonate mixtures at low concentrations,
$R_H\approx15$\,nm \cite{Park07}. At high concentrations, $R_H$ is
found to increase and stabilize at around 200-300\,nm, which results
in the decrease of the PVDF diffusivity \cite{Park04,Park07}. To keep
the formulation of the problem tractable, we do not account for
dependence of the diffusivity on the concentration of PVDF and keep
the value of the diffusion coefficient fixed at $D
=1.14\cdot10^{-10}$\,m$^2$/s.

\begin{table}
	\begin{center}
		\begin{tabular}{ |c|c|c|c|}
			\hline
			\hline
			Material              &  $\rho$ (g/cm$^3$)  & $\omega^{0}$ & $\omega^{\text{max}}$ \\
			\hline
			Solvent (NMP)         & 1.03                & 0.526          &  0     \\ 
			Polymer binder (PVDF) & 1.76                & 0.026          &  0.055 \\ 
			Graphite particles    & 2.21                & 0.449          &  0.945 \\ 
			\hline
			\hline
		\end{tabular}
		\caption{Typical densities $\rho$ for electrode film components and
			initial and final mass fractions, respectively, $\omega^{0}$ and
			$\omega^{\text{max}}$, for the electrode components used in Jaiser
			{\it et al} \cite{Jai16,Jai17b}. As noted previously, we do not
			account for the contribution of carbon black into our model, thus
			its mass fraction ($w_{cb} = 0.014$ \cite{Jai16,Jai17b}) has been
			added to the mass fraction of graphite particles. }
		\label{table1}
	\end{center}
\end{table}

\section{\label{results}Results and discussion}

In this section we first present and contrast typical model solutions
for both high and low drying rates. Then, we reproduce the
experimental procedure followed in \cite{Jai16} and demonstrate good
agreement between model solutions and experimental results. Finally,
we examine the effects of allowing time-dependent drying rates and
consider how this can be used to devise possible strategies to
optimize the drying process.

\subsection{Low and high drying rate limits}

The evolution of the binder distribution in a stable colloid is
determined by solving \eqref{eq_bin}-\eqref{init_bin} where the phase
volume fractions and fluxes are given by (\ref{mmaatt}). Although no
exact solutions to this problem are available, we can solve the
problem approximately in two different ways: (i) using matched
asymptotic expansions valid for limiting values of the Peclet number
$Pe = \gamma H_0/D$ (measuring the relative strength of advection to
diffusive transport), and; (ii) using a numerical scheme based on the
finite differences. We will contrast the two distinct limiting cases
where $Pe\ll1$ or $Pe\gg1$ which we will henceforth refer to as the
low drying rate (LDR) or high drying rate (HDR) case, respectively.
Details on the derivation of the asymptotic solutions and the
numerical scheme can be found in the supplementary information.

In the LDR limit ($Pe\ll1$) the concentration of binder is well
approximated by
\begin{align}\label{asym_LDR_2}
\frac{c(z,t)}{c_{0}} \approx \frac{(1-\phi_{\text{s}}^0)}{(\frac{H}{H_0}-\phi_{\text{s}}^0)} + 
Pe
\frac{(1-\phi_{\text{s}}^0)}{(\frac{H}{H_0}-\phi_{\text{s}}^0)^2\phi_{\text{l}}(t)^{1/2}} \left(\frac{z^2}{2H_0^2}-\frac{H^2}{6H_0^2}\right)\,.
\end{align}
where $H(t) = H_0 -\gamma t$ (obtained after integrating
\eqref{shsh}). In the HDR limit the concentration of binder is
approximately given by
\begin{align}\label{asym_HDR_2}
\frac{c(z,t)}{c_{0}} \approx 1 + A(t) \exp 
\left( -\frac{Pe (H-z)}{H_0\phi_{\text{l}}(t)^{3/2}} \right)\,, 
\end{align}
where now the time-dependent constant of integration $A(t)$ is the
solution of the following initial-value problem
\begin{align}\label{ode_A}
\dot{A} = -\frac{5}{2}A\frac{\dot{\phi}_{\text{l}}(t)}{\phi_{\text{l}}(t)} + 
\frac{Pe}{\phi_{\text{l}}(t)^{3/2}}\,, \qquad A(0) = 1\,.
\end{align}

\begin{figure}
	\centering
	\includegraphics[width=0.5\textwidth]{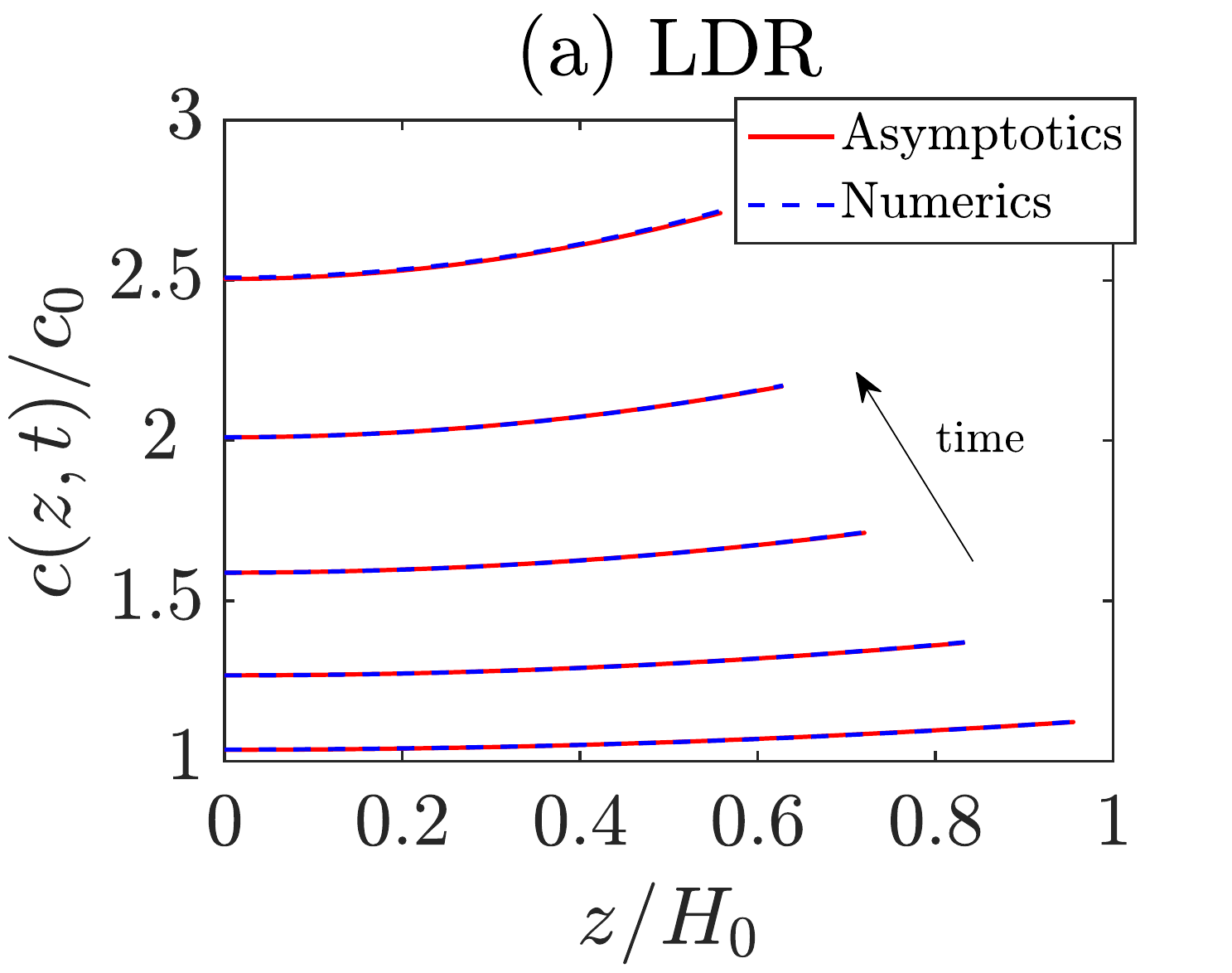}\includegraphics[width=0.5\textwidth]{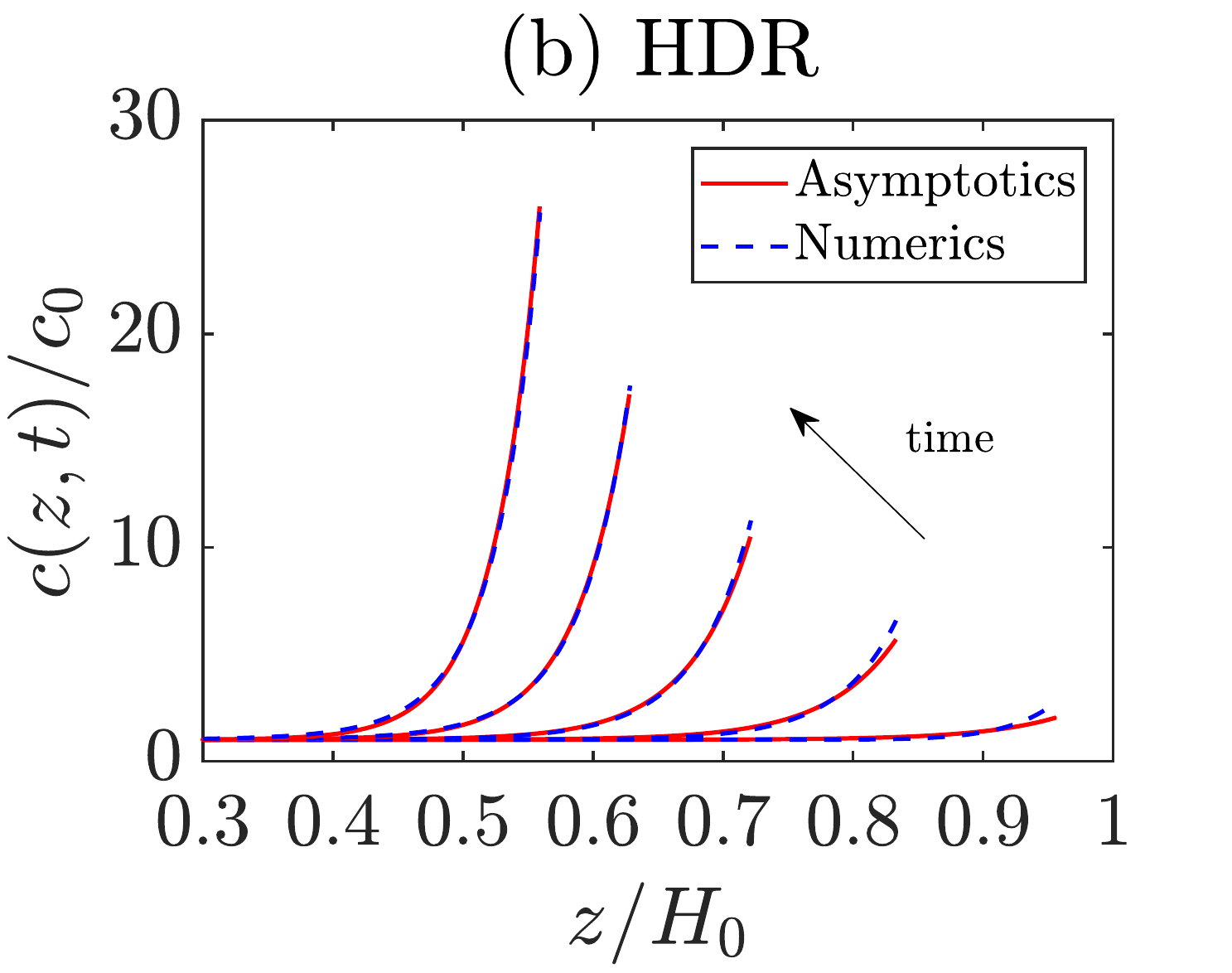}
	\caption{Numerical and asymptotic solutions of the binder transport
		model \eqref{eq_bin}-\eqref{init_bin}. Panel (a) shows concentration
		profiles corresponding to a low drying rate ($\gamma =
		1.25\cdot10^{-7}$\,m\,s$^{-1}$) for five values of time between
		39.8\,s and 401.92\,s. Panel (b) shows concentration profiles
		corresponding to a high drying rate ($\gamma =
		1.25\cdot10^{-5}$\,m\,s$^{-1}$) for five values of time between
		0.4\,s and 4.02\,s. The last profile in (a) and (b) corresponds to
		the concentration profile at the end of drying $t=t_{\text{end}}$.
		Note that panel (b) has been truncated at $z = 0.3$ for ease of
		viewing; the solution for $z<0.3$ is essentially flat.  }
	\label{fig1}
\end{figure}

In Figure \ref{fig1} we present typical solutions for both a low and
high drying rate by taking $\gamma = 1.25\cdot10^{-7}$\,m\,s$^{-1}$
($Pe \approx 0.1$) and $\gamma = 1.25\cdot10^{-5}$\,m\,s$^{-1}$ ($Pe
\approx 10$), respectively. The red solid lines correspond to the
asymptotic solutions \eqref{asym_LDR_2}--\eqref{ode_A} and blue dashed
lines to numerical solutions (see the supplementary information for
details). The two solution approaches exhibit very favourable
agreement despite the moderate sizes of the Peclet number used in each
case thereby validating both approaches. In panel (a) (low drying
rate) the concentration of binder progressively increases as solvent
evaporates with the distribution remaining almost homogeneous
throughout the whole drying process. For low drying rates ($Pe \ll 1$)
the drying time is relatively large and the velocity of the solvent
(upward) relatively small. Thus, advection is only able to induce
small gradients in the binder concentration and the diffusive process
has a long time to act to smooth out these gradients. The final binder
distribution is therefore relatively uniform. This is reflected in the
asymptotic solution \eqref{asym_LDR_2} where the leading order term
(and most dominant) is a function of time only and the dependence in
$z$ introduced only at the next order. Contrastingly, in Figure
\ref{fig1}(b) (high drying rate), the drying time is relatively small
and the velocity of the solvent relatively large. Here, diffusion has
less time to dissipate concentration gradients induced by solvent
advection. As a result, binder accumulates near the top surface of the
electrode film, which is captured by the exponential term in
\eqref{asym_HDR_2}. We can therefore conclude that low drying rates lead to a favourable homogeneous binder
profiles across the electrode film, whereas high drying rates tend to
unfavourably accumulate the binder near the evaporation top surface.

\subsection{Agreement with experiment}

We now utilise the model to reproduce and elucidate the experimental
results obtained in Jaiser et al \cite{Jai16}. In their work electrode
films were first dried at a high drying rate for a given period of
time $[0,t_{\text{trans}}]$. The drying rate was then decreased and
the drying process continued until the film was completely dry. The
main result in their study was a plot of $c_{\text{top}}=c|_{z=H,t=t_{end}}$ and
$c_{\text{bot}}=c|_{z=0,t=t_{end}}$ against $t_{\text{trans}}$, \ie the binder
concentrations at the top (evaporation surface) and bottom (current
collector) of the electrode at the end of the drying process
$t=t_{\text{end}}$. They observed that: (i) $c_{\text{top}}$ and
$c_{\text{bot}}$ are almost constant for sufficiently small
$t_{\text{trans}}$, (ii) there is then a small range of values of
$t_{\text{trans}}$ where $c_{\text{top}}$ increases whereas
$c_{\text{bot}}$ decreases beyond which, (iii) $c_{\text{top}}$ and
$c_{\text{bot}}$ once again saturate to constant values. We reproduce
this protocol in our model by taking the time-dependent drying rate used in \cite{Jai16}, namely
\begin{align}\label{asym}
\gamma(t) =
\begin{dcases}
1.16\,\mu\mbox{m}\,\mbox{s}^{-1} \quad (\mbox{Pe} = 0.94) \quad \mbox{for} \quad t<t_{\text{trans}}, \\
0.51\,\mu\mbox{m}\,\mbox{s}^{-1} \quad (\mbox{Pe} = 0.41) \quad \mbox{for} \quad t>t_{\text{trans}}.
\end{dcases} 
\end{align}

\begin{figure}
	\centering
	\subfigure[]{\includegraphics[width=0.50\textwidth]{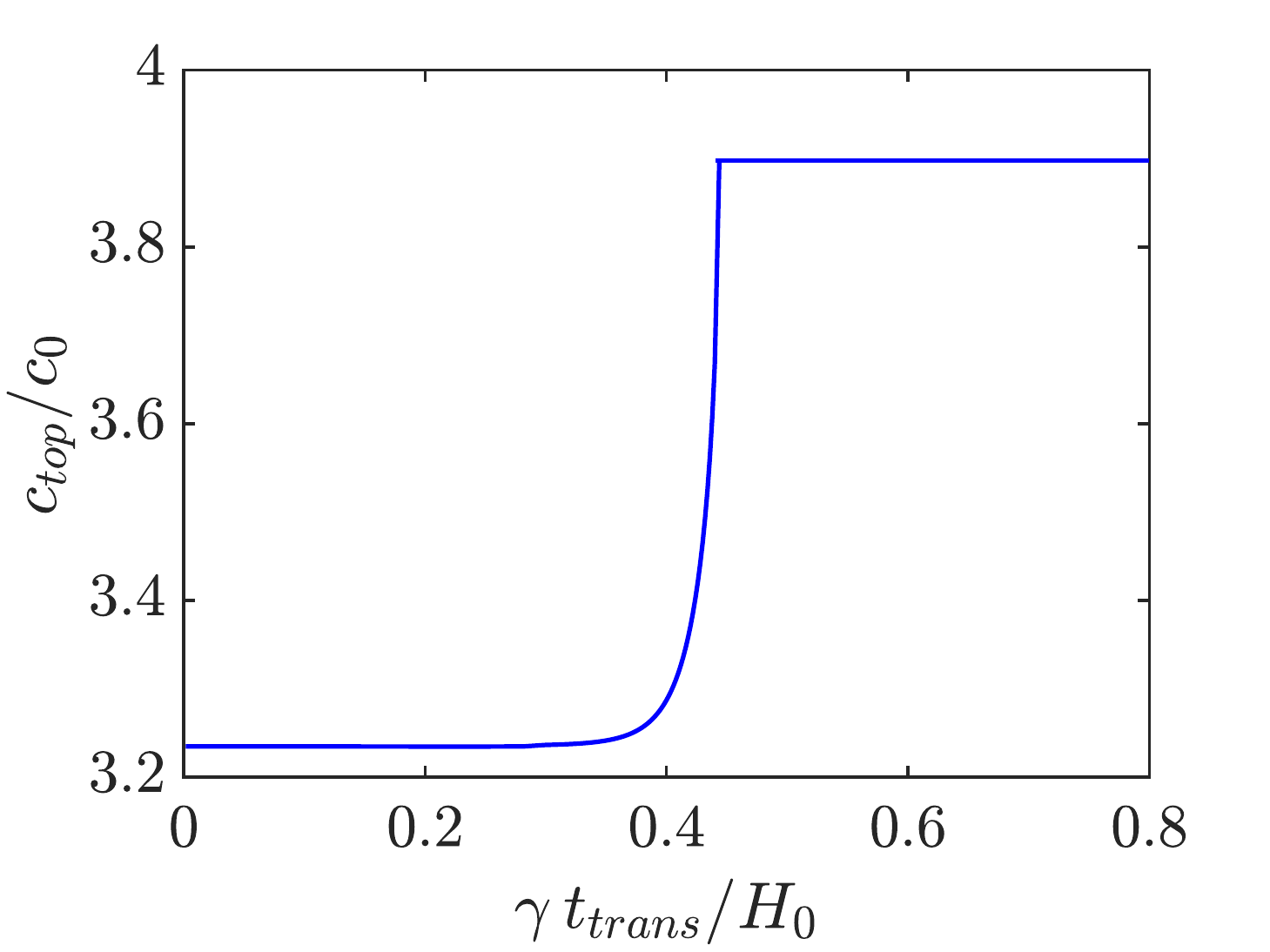}}\subfigure[]{\includegraphics[width=0.50\textwidth]{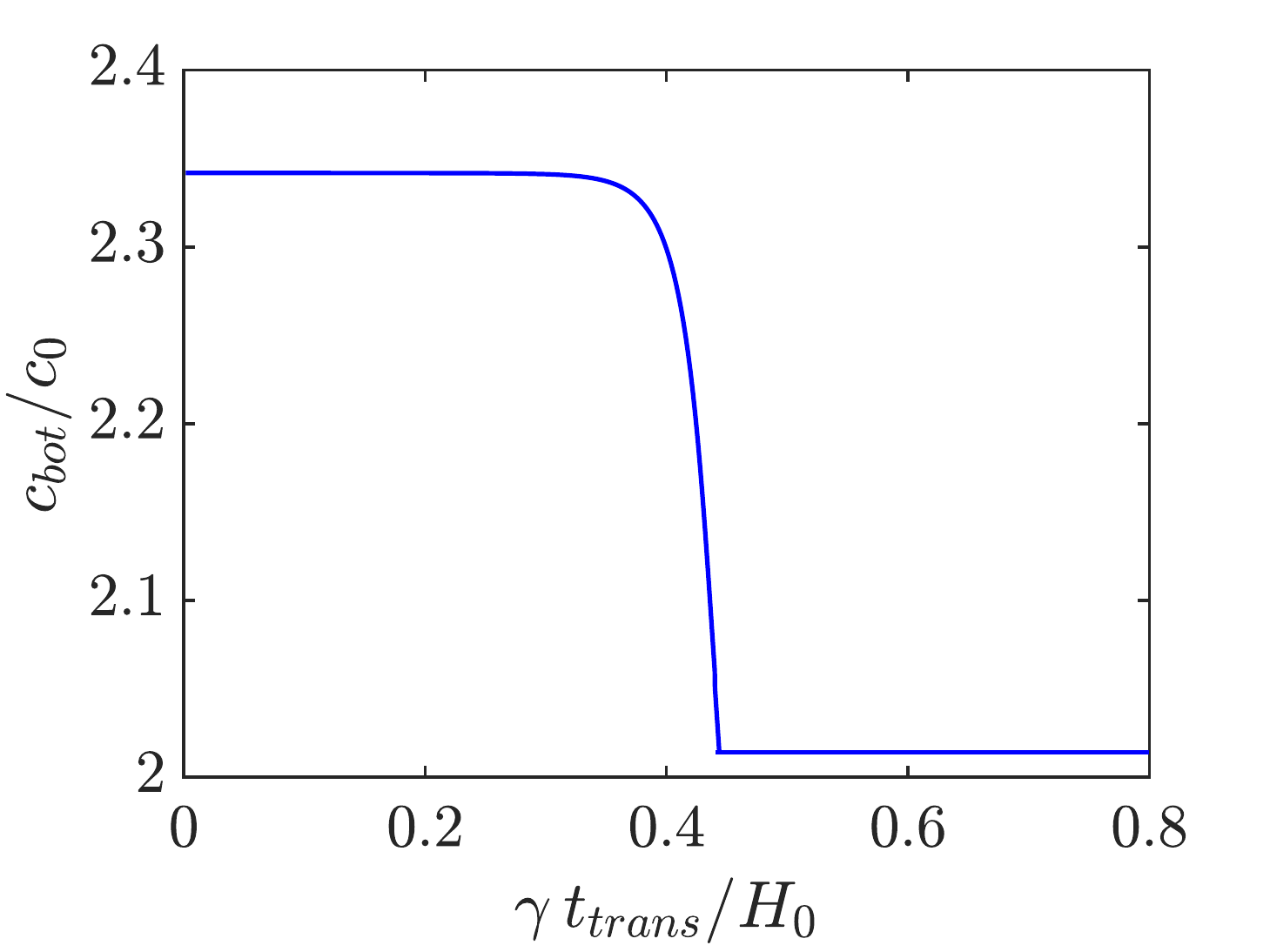}}
	\caption{Normalized concentration of binder on top and bottom of the
		electrode at the end of drying as a function of the nondimensional
		transition time between high and low drying rate. }
	\label{fig3}
\end{figure}

Figure \ref{fig3} shows the values of $c_{\text{top}}$ and
$c_{\text{bot}}$ for different choices of $t_{\text{trans}}$ as
determined using the numerical procedure described in the
supplementary information.  The plots show how $c_{\text{top}}$
increases slowly until $t_{\text{trans}}\gamma/H_0\approx 0.34$ and
then increases more rapidly until $t_{\text{trans}}\gamma/H_0 \approx
0.44$. For even larger values of $t_{\text{trans}}$, the decrease in
drying rate does not occur until after the electrode is completely
dry, thus $c_{\text{top}}$ remains constant. The evolution of
$c_{\text{bot}}$ behaves in an opposite fashion: it first decreases
slowly until $t_{\text{trans}}\gamma/H_0\approx 0.34$, then decreases
fast until $t_{\text{trans}}\gamma/H_0 = t_{\text{end}}\gamma/H_0
\approx 0.44 $ and stays constant for
$t_{\text{trans}}>t_{\text{end}}$. These results show strong
qualitative agreement with those presented in Figure 6 from
\cite{Jai16}.

\subsection{\label{viable}Identifying viable constant drying rates}

\begin{figure}
	\centering
	\includegraphics[width=0.5\textwidth]{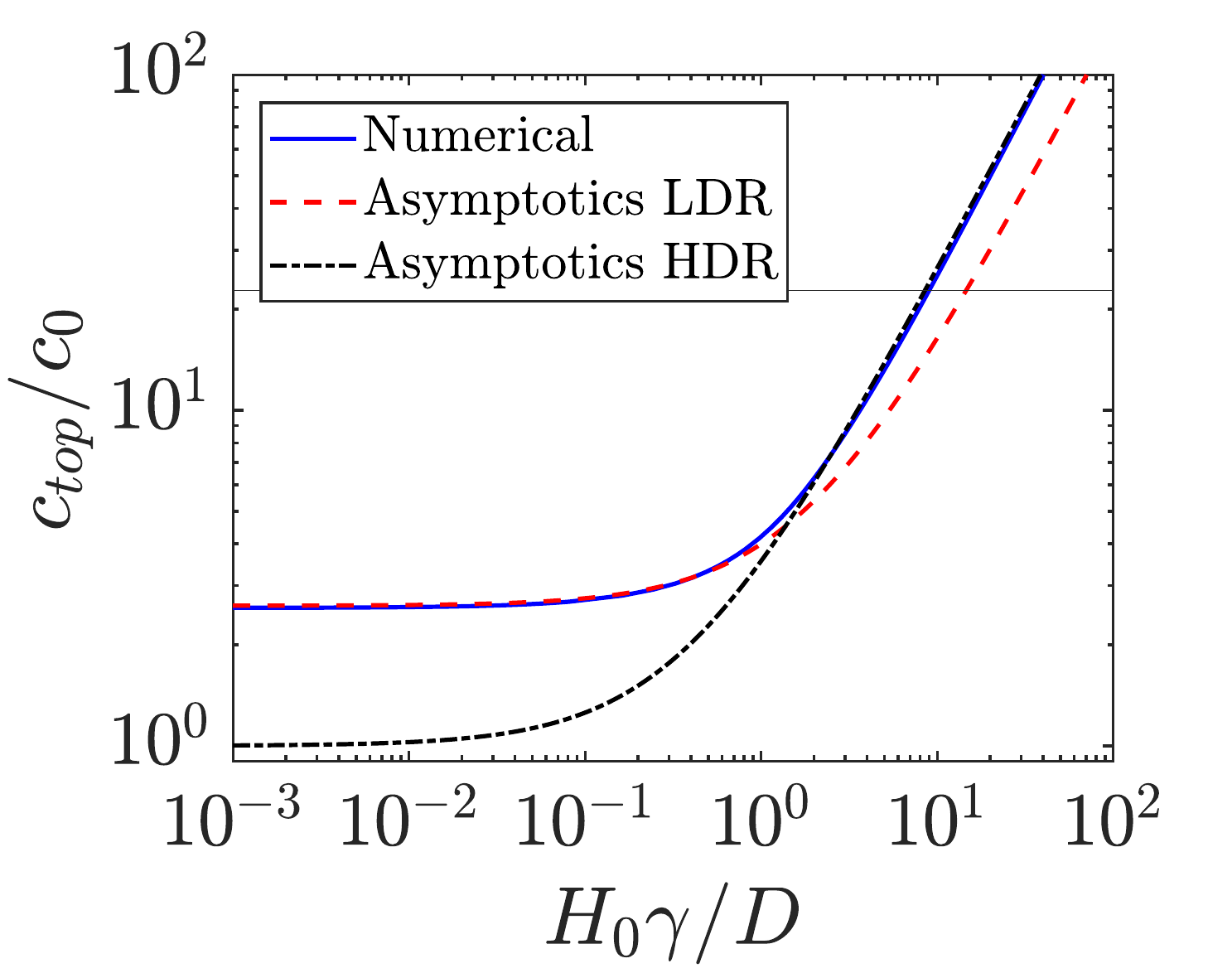}
	\caption{Binder concentration on top of the electrode at the end of
		drying as a function of the drying rate $\gamma$. The solid line
		corresponds to the numerical solution and the the dashed and
		dash-dotted lines correspond to the asymptotic solution for low and
		high drying rate, respectively. The horizontal line denotes the
		binder concentration at which crystallisation begins to occur at
		$T=60$\,$^o$C, corresponding to $c(z,t)/c_{0} = 22.75$
		\cite{Ma2011}, which represents an orientative upper bound for the
		concentration in our model.}
	\label{fig2}
\end{figure}

We now use the model to identify the largest rate at which an
electrode can afford to be dried without inducing unacceptably large
gradients in the binder concentration. In Figure \ref{fig2} we present
the concentration of binder on top of the electrode at the end of
drying as a function of the Peclet number. The value of the drying
rate by which concentration gradients remain relatively small (\ie
$c_{\text{top}}$ does not increase substantially) corresponds to $Pe =
\gamma H_0/D \lessapprox 1$. For the values of $H_0$ and $D$ estimated
in \S\ref{sec_param} this yields $\gamma \lessapprox 1\cdot10^{-6}$\,m/s.

From Figure \ref{fig2} we also see that, unless a very aggressive
drying rate is used, it seems unlikely that crystallization of the PVDF
will begin to occur until after the film is fully consolidated, the
point at which our model is no longer valid and our simulations are
terminated.

\subsection{Exploiting variable drying rates}

\begin{figure}
	\centering
	\includegraphics[width=0.5\textwidth]{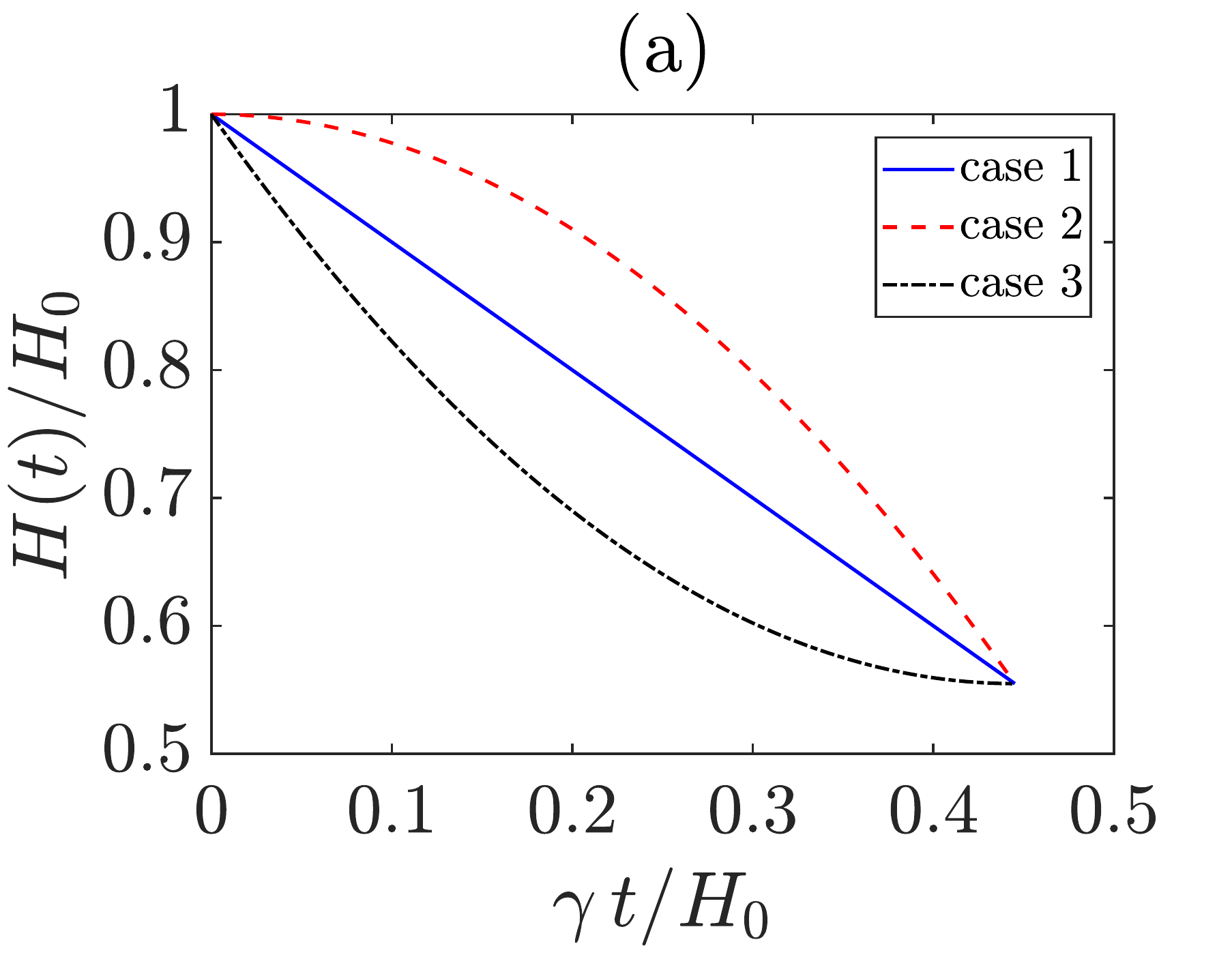}\includegraphics[width=0.5\textwidth]{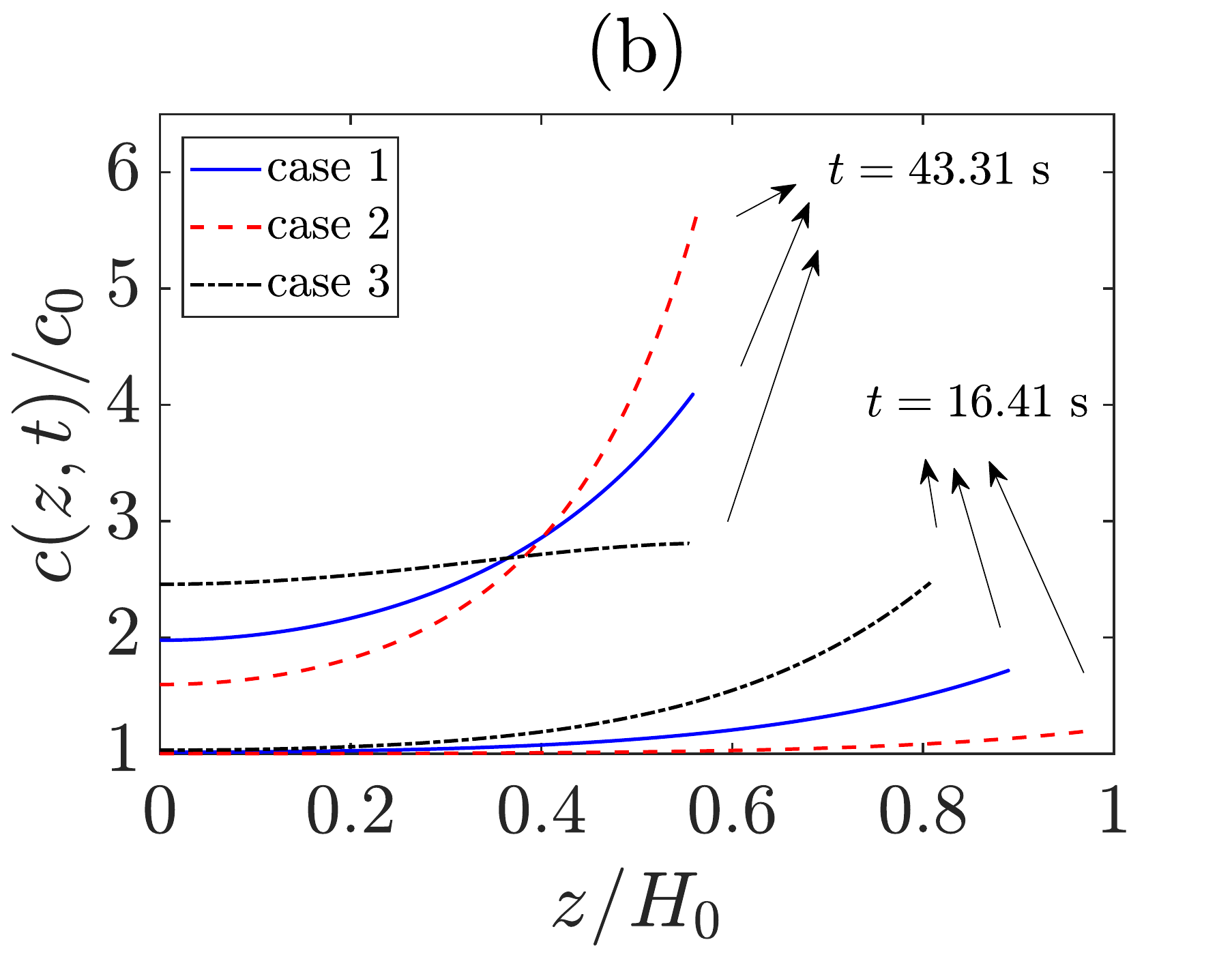}
	\caption{(a) Evolution of the position of the top surface for the
		three drying rates considered, cf.~\eqref{case1}--\eqref{case3}. (b)
		Concentration of binder near the beginning ($t=16.41$\,s) and at the
		end ($t=43.31$\,s) of the drying process for each case. }
	\label{fig4}
\end{figure}

As we have demonstrated, the drying process should be carried out
slowly to prevent an undesirable accumulation of binder near the
evaporation surface. However, from an industrial point of view, short
drying processes are preferred in order to increase throughput
\cite{WoodIII2015234}. The model is now used to investigate whether
(and to what extent) allowing time-dependent drying rates can be
helpful in simultaneously achieving more homogeneous binder
distributions and shorter drying times.  To do so, we consider three
different drying protocols: (Case 1) a constant drying rate, (Case 2)
a linearly increasing drying rate, and (Case 3) a linearly decreasing
drying rate, as outlined below:
\begin{align}
\text{Case 1:} && \gamma(t) &= \gamma_0 & \mbox{for} \quad 0 \leq t \leq t_{end}, \label{case1}\\
\text{Case 2:} && \gamma(t) &= 2 \gamma_0 \frac{t}{t_{\text{end}}} & \mbox{for} \quad 0 \leq t \leq t_{end}, \label{case2}\\
\text{Case 3:} && \gamma(t) &= 2 \gamma_0 \left(1-\frac{t}{t_{\text{end}}}\right) & \mbox{for} \quad 0 \leq t \leq t_{end}, \label{case3}
\end{align}
and select $\gamma_0 =1.16$\,$\mu\text{m}$\,s$^{-1}$. Note that in
defining \eqref{case1}--\eqref{case3} we have ensured that the time
taken to fully consolidate the film, $t_{\text{end}}$, is the same in
all three cases (see supplementary information). The evolution of the
position of the evaporating surface is represented in Figure
\ref{fig4}a. In Figure \ref{fig4}b we show the concentration of binder
across the electrode at two different times during the drying process
(near the beginning and at the end) for each choice of the drying
rate.

We observe that choosing a linearly decreasing drying rate (case 3)
gives the most evenly distributed binder concentration whilst choosing
a linearly increasing one yields the worst results. This suggests
that, if the goal is obtaining an acceptably homogeneous distribution
of binder and a short drying time, the best procedure is to dry the
electrode at a high drying rate at the beginning and at a low drying
rate near the end of the process. This can be rationalised by noting
that even though large binder concentration gradients may be
established by the initial high drying rate, so long as the rate is
dropped towards the end of the process, diffusive effects overcome the
upward convection of the solvent and have sufficient time to act to
dissipate these inhomogeneities. These results, which have been
obtained for a fixed drying time, can be interpreted in the sense of
minimizing the drying time: approximately the same concentration of
binder as obtained at a given constant drying rate could be achieved
in a shorter time by using HDR at the beginning and LDR near the end.
An interesting and open question concerns whether the drying process
can be further optimised by allowing a more complicated time-dependent
behaviour for $\gamma(t)$. Work to address this question is already
underway and will be reported in a future study.

\section{\label{conc} Conclusions} 

We have presented a mathematical model that predicts the mass
transport and evolution of binder concentration during the drying of
lithium-ion battery electrodes. We have found that higher drying rates
tend to induce larger binder concentration gradients because to a
combination of: (i) the more aggressive evaporation rates causing a
larger (upward) convection of the binder solvent and; (ii) the
decreasing drying time allows less opportunity for diffusion to
redistribute the binder evenly throughout the film. We have
demonstrated that the model satisfactorily reproduces recently
published experimental results of binder migration phenomena during
drying. Finally, we have shown that a sound strategy to reduce the
drying time whilst simultaneously maintaining small variations in the
binder concentration is to initially apply a period of high drying
rate and to then decrease this rate towards the end of the process.

\section*{Acknowledgments}

We are thankful to G.~Goward, I.~Halalay, X. Huang, M. Jiang, K. J. Harris, M. Z. Tessaro, H. Liu and
S.~Schougaard for helpful discussions. Funding for this research was
provided by the Natural Sciences and Engineering Research Council of
Canada (Collaborative Research \& Development grant CRDPJ 494074) and
by General Motors of Canada.

\bibliographystyle{plain}
\bibliography{references}

\newpage
\appendix

\section{Supplementary Information}

\subsection{Nondimensional model of binder transport}

We introduce the dimensionless variables 
\begin{align}\label{dimless}
\hat{c} = \frac{c}{c_{0}}\,, \quad 
\hat{z} = \frac{z}{H_0}\,, \quad \hat{H} = \frac{H}{H_0}\,,\quad \hat{t} = \frac{\gamma}{H_0}t\,
\end{align}
into the binder model (14)--(16). Dropping the `` $\hat{ }$ ''
symbols, the governing equation for the concentration of binder
becomes
\begin{align}\label{mod2_eq3_non}
{Pe}~\left[\phi_{\text{l}}(t)\frac{\partial c}{\partial t} -\dot{\phi}_l(t) z \frac{\partial c}{\partial z}\right] = \phi_{\text{l}}(t)^{3/2}\frac{\partial^2 c}{\partial z^2}\,,
\end{align}
where $\phi_{l}(t)$ is given by Eq.~(13) in the main text. The
dimensionless parameter ${Pe} = \gamma H_0 /D_{0}$ is the Peclet
number measuring the relative strength of binder advection due the
upward transport of solvent versus the diffusive transport of binder.
If $\gamma$ is small then diffusion dominates and ${Pe}\ll1$, whereas
if $\gamma$ is large then evaporation dominates over diffusion and
${Pe}\gg1$. In the next section, we will seek approximate solutions
for these two distinct regimes. Using the rescalings defined in
\eqref{dimless} the boundary conditions (15) and the initial
condition (16) become
\begin{align}\label{mod2_eq4_non}
\left.\frac{\partial c}{\partial z}\right|_{z = 0} = 0 \,, \qquad 
\phi_{\text{l}}(t)^{3/2}\left.\frac{\partial c}{\partial z}\right|_{z = H(t)} =  {Pe}\left.c\right|_{z = H(t)}\,, \qquad c(z,0) = 1\,.
\end{align}
The position of the evaporating surface moves according to
\begin{align}\label{mov}
H(t) = 1-t\,
\end{align}
which is obtained integrating the dimensionless version of Eq.~(11)
from the main text, $\dot{H}=-1$, and applying the initial condition
$H(0)=1$.

Binder does not enter or leave the electrode film during the drying
process, therefore mass is conserved and the following condition must
be satisfied at all times
\begin{equation}\label{cons} 
\frac{d}{dt} \left( \int_0^{H(t)} \phi_{\text{l}} c\,  dz \right) =0\,.
\end{equation}
This equation can be integrated using the initial conditions
$c(z,0)=1$ and $H(0)=1$ to give
\begin{align}\label{mod2_cons2_non}
\int_{0}^{H} \phi_{\text{l}}(t) c \, dz = 1-\phi_{\text{s}}^{0}\,.  
\end{align}
Expression \eqref{mod2_cons2_non} will be used in the derivation of
the asymptotic solutions in the sections below.

\subsection{Asymptotic solution for low drying rate}
\label{asym_sec}

We now seek an approximate solution for small value of $Pe$ (low drying 
rate). It is clear (and can easily be shown from a balance at leading order) that an appropriate expansion is of the form 
\begin{align}
c = c_{(0)} + {Pe}~c_{(1)} + \mathcal{O}({Pe}^2).
\end{align} 
Using it in \eqref{mod2_eq3_non}--\eqref{mod2_cons2_non} we obtain the
leading-order problem 
\begin{align}\label{mod2_LDR_1}
\frac{\partial^2 c_{(0)}}{\partial z^2} = 0\,,\qquad \left.\frac{\partial c_{(0)}}{\partial z}\right|_{z = 0} = 0\,,\qquad \left. \frac{\partial c_{(0)}}{\partial z}\right|_{z = H} = 0\,. 
\end{align}
This has solution $c_{(0)}=c_{(0)}(t)$, and in order to simeltaneously satisfy \eqref{mod2_cons2_non} and \eqref{mod2_eq4_non}
\begin{align}
c_{(0)} = \frac{1-\phi_{\text{s}}^0}{H(t)-\phi_{\text{s}}^0}\,.  
\end{align}
The first-order problem is then
\begin{subequations}
	\label{mod2_LDR_ord1}
	\begin{align}
	& \phi_{\text{l}}(t)\frac{\partial c_{(0)}}{\partial t} -\dot{\phi}_l(t) z \frac{\partial c_{(0)}}{\partial z} = \phi_{\text{l}}(t)^{3/2}\frac{\partial^2 c_{(1)}}{\partial z^2}\, \label{mod2_LDR_ord1a} \\
	& \left. \frac{\partial c_{(1)}}{\partial z}\right|_{z = 0} = 0\,, 
	\qquad \left. c_{(0)}\right|_{z = H} = \phi_{\text{l}}^{3/2}\left.\frac{\partial c_{(1)}}{\partial z}\right|_{z = H}\,. 
	\end{align}
\end{subequations}
We integrate the PDE \eqref{mod2_LDR_ord1a} and apply the boundary
condition at $z=0$. Then, since the boundary condition at $z=H$ does
not provide additional information, we use the relation/solvability condition
\eqref{mod2_cons2_non} to fully determine the solution of
\eqref{mod2_LDR_ord1}. We thus obtain
\begin{align}
c_{(1)} = \frac{\dot{c}_{(0)}}{\phi_{\text{l}}^{1/2}}\left(\frac{z^2}{2}-\frac{H^2}{6}\right)
\end{align}
and finally
\begin{align}\label{asym_LDR_SEI}
c = c_{(0)} + {Pe}~\frac{\dot{c}_{(0)}}{\phi_{\text{l}}^{1/2}}\left(\frac{z^2}{2}-\frac{H^2}{6}\right)+ \mathcal{O}({Pe}^2).
\end{align}
Replacing the dimensionless variables in \eqref{asym_LDR_SEI} with the
corresponding dimensional ones we obtain expression (21) from the main
text.

\subsection{Asymptotic solution for high drying rate} 

The high drying rate limit corresponds to ${Pe}\rightarrow\infty$
which suggests the introduction of a new small parameter $\varepsilon
= 1/{Pe}$. In this case, the governing equation, \eqref{mod2_eq3_non}, becomes
\begin{align}\label{mod2_HDR_asym1}
\phi_{\text{l}}(t)\frac{\partial c}{\partial t} -\dot{\phi}_l(t) \frac{\partial c}{\partial z} = \varepsilon \phi_{\text{l}}(t)^{3/2}\frac{\partial^2 c}{\partial z^2} \qquad \text{on} \ 0 < z < H(t)\, 
\end{align}
and the boundary condition at $z = H(t)$ takes the form
\begin{align}\label{mod2_HDR_asym2}
\varepsilon \phi_{\text{l}}(t)^{3/2}\left.\frac{\partial c}{\partial z}\right|_{z = H(t)} =   \left.c\right|_{z = H(t)}\,. 
\end{align}
It is clear that the problem
\eqref{mod2_HDR_asym1}--\eqref{mod2_HDR_asym2} is of a
singular-perturbation type as the term with the highest-order
derivative vanishes in the limit $\varepsilon\rightarrow0$, which
reveals the presence of a boundary layer near $z = H(t)$. In what
follows we therefore distinguish two regions: the bulk ((I)), formed
by most of the electrode, and the boundary layer ((II)), the small
region near the boundary $z=H(t)$. We start by analysing the bulk
region.

\textbf{The bulk (I):} In the bulk region we expand the solution as
follows
\begin{align}
c(z,t) = c_{(0)}^{(I)} + \mathcal{O}(\varepsilon).
\end{align} 
Inserting the above into \eqref{mod2_HDR_asym1} reveals that the
leading-order problem in the bulk is
\begin{align}\label{}
\phi_{\text{l}}(t)\frac{\partial c_{(0)}^{(I)}}{\partial t} -\dot{\phi}_l(t) \frac{\partial c_{(0)}^{(I)}}{\partial z} = 0\,. 
\end{align}
It is clear that a solution to this equation satisfying the initial
condition and all boundary conditions except for
\eqref{mod2_HDR_asym2} is
\begin{equation}
c^{(I)}_{(0)} = 1.
\end{equation}
We now proceed to analyse the boundary layer near $z = H$ to find a
solution that satisfies \eqref{mod2_HDR_asym2}.

\textbf{Boundary layer near the evaporation surface (II):} To examine
the behaviour of solutions here we rescale the spatial coordinate as
follows
\begin{equation}
\varepsilon y = H(t) - z, 
\end{equation}
transforming \eqref{mod2_HDR_asym1}--\eqref{mod2_HDR_asym2} into
\begin{equation}
\begin{aligned}
& \varepsilon\left(\phi_{\text{l}} \frac{\partial c^{(II)}}{\partial t} - \dot{\phi_{l}} y \frac{\partial c^{(II)}}{\partial y}\right) = \frac{\partial c^{(II)}}{\partial y} + \phi_{\text{l}}^{3/2}\frac{\partial^2 c^{(II)}}{\partial y^2} \,, \\
& \phi_{\text{l}}(t)^{3/2}\left.\frac{\partial c^{(II)}}{\partial y}\right|_{y = 0} =   \left.c^{(II)}\right|_{y = 0}\,. 
\end{aligned}
\end{equation}
We then expand the solution as follows
\begin{align}
c^{(II)} = c^{(II)}_{(0)} + \varepsilon c^{(II)}_{(1)} + \mathcal{O}(\varepsilon^2)\,
\end{align}
which provides the leading and first-order problems
\begin{alignat*}{2}
&\mathcal{O}(1): & \qquad
0 & = \frac{\partial c^{(II)}_{(0)}}{\partial y} + \phi_{\text{l}}^{3/2}\frac{\partial^2 c^{(II)}_{(0)} 
}{\partial y^2}\,, \\ 
&& \left.c^{(II)}_{(0)}\right|_{y=0} & = -\phi_{\text{l}}^{3/2} \left.\frac{\partial c^{(II)}_{(0)}}{\partial y}\right|_{y=0} \,, \\
& \mathcal{O}(\varepsilon):&  \qquad\qquad 
\phi_{\text{l}} \frac{\partial c^{(II)}_{(0)}}{\partial t} - \dot{\phi_{l}} y \frac{\partial c^{(II)}_{(0)}}{\partial y}
& = \frac{\partial c^{(II)}_{(1)}}{\partial y} + \phi_{\text{l}}^{3/2} \frac{\partial^2 c^{(II)}_{(1)}}{\partial y^2}\,, \\ 
&& \left.c^{(II)}_{(1)}\right|_{y=0} & = -\phi_{\text{l}}^{3/2} \left.\frac{\partial c^{(II)}_{(1)}}{\partial y}\right|_{y=0} \,. 
\end{alignat*}
Solving the above problems we obtain 
\begin{align}
c_{(0)}^{(II)} &= A(t)e^{-y/\phi_{\text{l}}^{3/2}}\,, \\
c_{(1)}^{(II)} &= (p_1 + p_2 \phi^{3/2}_{\text{l}})\left\lbrace \phi^{3/2}_{\text{l}} -e^{-y/\phi_{\text{l}}^{3/2}}\left[ \phi^{3/2}_{\text{l}} +y +\frac{(p_2y^2+2\phi_{\text{l}}^{3/2}B(t))}{2(p_1 + p_2 \phi^{3/2}_{\text{l}})}\right]  \right\rbrace \,,
\end{align}
where 
\begin{align}
p_1 = \phi_{\text{l}}\dot{A}\,, \qquad  p_2 = \frac{5}{2}
\frac{\dot{\phi}_{\text{l}}}{\phi_{\text{l}}^{3/2}}A\,
\end{align}
and $B(t)$ is a constant of integration that we leave undetermined.
Matching to the outer region
\begin{align}\label{match}
\left.(c^{(II)}_{(0)} + \varepsilon c^{(II)}_{(1)})\right|_{y \rightarrow \infty} \sim 
\left.c^{(I)}_{(0)}\right|_{z\rightarrow H(t)}\,,
\end{align}
provides the following initial value problem for $A(t)$
\begin{align}\label{ode_A}
\dot{A} = -\frac{5}{2}\frac{\dot{\phi}_l}{\phi_{\text{l}}}A + \frac{1}{\varepsilon\phi_{\text{l}}^{3/2}}\,, \qquad A(0) = 1\,. 
\end{align}
Now, by adding the inner solution $c^{(II)}\approx c_{(0)}^{(II)}$ and
the outer solution $c^{(I)}\approx c_{(0)}^{(I)}$, we develop the
uniformly valid approximation
\begin{align}\label{asym_HDR_2}
c_{\text{uni}} = c^{(I)} + c^{(II)} = 1 + A(t) e^{-\frac{(H-z)}{\varepsilon \phi_{\text{l}}^{3/2}}}\,, 
\end{align}
where the value of $A(t)$ is obtained by numerically integrating
\eqref{ode_A} using the Matlab function \texttt{ode45}.

Replacing the dimensionless variables in \eqref{asym_HDR_2} with the
dimensional ones we obtain expression (22) from the main text.


\subsection{Numerical solution} 

Coordinate transformations mapping domains with variable boundaries to
fixed domains are widely used when computing numerical solutions of
moving-boundary problems because of the advantage of working with
fixed domains. To solve our model numerically we follow this approach
and map the space variable $z$ in \eqref{mod2_eq3_non} to the unit
domain $[0,1]$ by means of the transformation $\xi=z/H(t)$. Then, the
governing equation becomes
\begin{align}\label{eq_num}
a(t) \frac{\partial u}{\partial t} &= b(\xi,t)\frac{\partial u}{\partial \xi} + 
\frac{\partial^2u}{\partial \xi^2} \quad\text{on}\ 0\leq\xi\leq1 \,,
\end{align}
where $u(\xi,t)$ represents the concentration of binder in the
transformed variable and the coefficients $a(t)$ and $b(\xi,t)$ take
the form
\begin{align}\label{a_b_mod2}
a(t) = {Pe}~\phi_{\text{l}}(t)^{-1/2}H^2\,, \qquad b(\xi,t) = - {Pe}~\phi_{\text{l}}(t)^{-3/2}H\dot{H}\xi\,. 
\end{align}
The corresponding boundary conditions are  
\begin{align}\label{bc_mod2_fixed}
\left. \frac{\partial u}{\partial \xi}\right|_{\xi = 0}= 0\,, \qquad \phi_{\text{l}}(t)^{3/2}  \left. \frac{\partial u}{\partial \xi}\right|_{\xi = 1}  = -{Pe}~H\dot{H} \left.c\right|_{\xi = 1}\,
\end{align}
and the initial condition is 
\begin{align}\label{ic_num}
u(\xi,0) = 1\,. 
\end{align}
To discretize problem \eqref{eq_num}--\eqref{ic_num} we use
second-order central differences in space and the Crank-Nicolson
scheme in time. We also employ one-sided second-order finite
differences to discretize the boundary conditions, thereby ensuring
that the solution is overall second-order accurate with respect to
discretisation of both the space and time variables. The numerical
approach is implemented in Matlab.

\subsection{Comparison with experiment: numerical procedure}

In order to reproduce the experimental approach from \cite{Jai16} we
use the following procedure. First, we solve the model using
$\gamma_{1}$ until $t = t_{\text{trans}}$. Then, we take
$c(z,t_{\text{trans}})$, $H(t_{\text{trans}})$ as the new initial
conditions and solve the problem with $\gamma_{2}$. We repeat the
process for increasing values of $t_{\text{trans}}$. To make the
procedure as close as possible to the actual experiments, we take the
parameter values from Jaiser {\it et al} \cite{Jai16}.  The mass flux
imposed at the electrode surface was switched at every
$t_{\text{trans}}$ from $q_{1} = 1.19$\,g\,m$^{-2}$\,s$^{-1}$ to
$q_{2} = 0.52$\,g\,m$^{-2}$\,s$^{-1}$. Then, dividing these values by
the density of NMP, we obtain the corresponding drying rates
$\gamma_{1} = 1.16$\,$\mu$m\,s$^{-1}$ ($Pe = 0.94$) and $\gamma_{2} =
0.51$\,$\mu$m\,s$^{-1}$ ($Pe = 0.41$).

The value of the diffusivity coefficient $D$ is chosen such that the
ratio $\min{(c_{\text{top}})}/\max{(c_{\text{top}})}$, where
$\min{(c_{\text{top}})}$ and $\max{(c_{\text{top}})}$ correspond to
the lower and upper bounds in Figure 3a, computed from our numerical
solution coincides with the corresponding ratio of the experimental
values $\min{(c_{\text{top}})}/\max{(c_{\text{top}})}\approx0.83$
obtained from Figure 6 (top panel) in Jaiser {\it et al} \cite{Jai16}.
We found that in order for
$\min{(c_{\text{top}})}/\max{(c_{\text{top}})}\approx0.83$  
the diffusion coefficient has to be $D \approx
1.14\cdot10^{-10}$\,m$^2$\,s$^{-1}$.  In addition, in such case we
also obtain the ratio $\min{(c_{\text{bot}})}/\max{(c_{\text{bot}})}
\approx 0.87$ which is very close to 0.86, the ratio obtained in
Jaiser {\it et al} \cite{Jai16} (bottom panel in Figure 6).

\subsection{Variable drying rate}

In defining the variable drying rates (25)--(27) we kept
$t_{\text{end}}$ constant by imposing the constraint
\begin{align}
\int_0^{t_{\text{end}}}\gamma(t)\,dt = \gamma_0\, t_{\text{end}}\,
\end{align} 
which ensures that the same amount of solvent is removed from the
slurry in each case. An expression for $t_{\text{end}}$ can be found
using relation (13)a together with the equation for the position of
the evaporation surface \eqref{mov} in dimensional form, $H(t)=H_0 -
\gamma_0 t$.  Noting that $\phi_{\text{s}}|_{t=t_{\text{end}}} =
\phi_{\text{s}}^{\text{max}}$, we have
\begin{align}
t_{\text{end}} = \frac{H_0}{\gamma_0}\frac{(\phi_{\text{s}}^{\text{max}}-\phi_{\text{s}}^{0})}{\phi_{\text{s}}^{\text{max}}}\,. 
\end{align}


\end{document}